\documentclass[pra,aps,twocolumn,nofootinbib,floatfix]{revtex4-2}
\usepackage{graphicx}
\usepackage{bm}
\usepackage{hyperref}
\usepackage{amsmath}

\newcommand{\cor}[1]{D^{(#1)}}
\def\be{\begin{equation}}
\def\ee{\end{equation}}
\def\ba{\begin{eqnarray}}
\def\ea{\end{eqnarray}}
\def\half{\frac{1}{2}}

\def\part{\partial}

\def\tr{\mbox{tr}}

\def\dag{\dagger}
\def\a{\alpha}
\def\b{\beta}
\def\gam{\gamma}
\def\Gam{\Gamma}

\def\D{\Delta}
\def\ome{\omega}

\def\lam{\lambda}

\def\eps{\epsilon}
\def\kap{\kappa}
\def\la{\langle}
\def\ra{\rangle}

\def\cN{{\cal N}}
\def\bin{b_{in}}
\def\bout{b_{out}}

\def\tbin{\tilde{b}_{in}}
\def\tbout{\tilde{b}_{out}}
\def\ta{\tilde{a}}
\def\tp{t_+}
\def\hC{\hat{C}}
\def\hL{\hat{L}}
\def\nth{n_{th}}

\def\Dt{\Delta t}

\def\dome{\delta \omega}
\def\DSout{\Delta S_{out}}
\def\kmax{k_{\max}}

\begin{document}

\title{Entropy flow in a parametric amplifier}
\author{Sergei Khlebnikov}
\affiliation{Department of Physics and Astronomy, Purdue University, West Lafayette, 
IN 47907, USA}


\begin{abstract}
Computations of entropy in thermodynamics rely on discreteness of the spectra of the subsystems.
We argue that, for cases with continuous spectra (typically, radiation), 
there is a useful definition
of entropy flow based on discretizing the signal into Gabor's ``atoms,'' say,
by means of a windowed Fourier transform. In particular, applying this method to 
a degenerate 
parametric amplifier (paramp) driven by a classical pump and coupled to a zero-temperature 
Markovian bath, we find that the output entropy flux vanishes at large times, 
even though the energy and photon number fluxes remain nonzero. This is consistent with the
manner in which the paramp is expected to release information about its initial state.
We relate the quenching of the entropy flow to development of the off-diagonal coherences in 
the output and discuss possible relevance of this mechanism to
the black-hole information problem. We also propose to use measurements of the off-diagonal 
coherences as a means of extracting the rates at which high-entropy subsystems release 
information to linear environments. 
\end{abstract}

\maketitle

\section{Introduction}
In statistical mechanics, there are various quantities that go under the name of 
entropy, each serving a somewhat different purpose. For a subsystem of a larger system,
one can define the entanglement entropy by the von Neumann formula 
$S = -\tr (\rho \ln \rho)$, where $\rho$ is the subsystem's density matrix. One can then
add up the entropies computed for the individual 
subsystems to form what is known as the entropy of the system in thermodynamics 
\cite[\S 7]{LL}. For an excited state of the system, it corresponds in many cases
to a finite entropy {\em density}.
The present work has been motivated by the question if a definition of
entropy {\em flux}
useful for analysis of quantum circuits can be obtained similarly, except that,
instead of partitioning
the system in space (as often done in thermodynamics), we partition the signal in time.
The definition being useful may mean, for instance, that it tracks correctly the information
flow in the circuit. 

A common method of discretization of classical signals is Gabor's ``atoms'' \cite{Gabor},
the minimal area regions in the time-frequency plane, 
as obtained for instance by means of a windowed Fourier transform. In the quantum case,
a windowed Fourier transform has been used in a study of squeezing in the output
of a laser \cite{modes_of_univ} and in computation of the photon number
flux on a transmission line \cite{Clerk&al}. Here, we apply the full set of Gabor's atoms
generated by a windowed cosine transform to obtain a definition of the entanglement 
entropy flux and compute that flux in a simple physical system.  

The system in question is a degenerate parametric amplifier 
(paramp) driven by a classical pump and coupled to a zero-temperature Markovian 
bath---in certain respects, a canonical system in the theory of quantum 
circuits (see, for example, Ref.~\cite{Collett&Gardiner}).
The paramp itself does not require any discretization, so its
entanglement entropy can be computed by standard methods. 
Adopting the framework of the input-output theory 
\cite{Collett&Gardiner,Gardiner&Collett}, we find that the entropy of the paramp
(operated below the instability threshold) reaches a constant asymptotic 
value at large times. By the well-known property of the entanglement entropy, this implies
that, if the entire system (the paramp plus the bath) starts out in a pure state,
the asymptotic entropy of the bath will also be a constant and equal to that of the paramp.
This, in turn, suggests that, after some initial transient, the entropy flux between the 
two subsystems should cease, even though the energy and photon number fluxes may be
expected to continue.

To look at this effect in more detail, we compute the output entropy flux by discretizing 
the output signal via a windowed cosine transform. At large times, the resulting 
discrete system is Gaussian, and we compute its entanglement entropy numerically  
using symplectic diagonalization \cite{Williamson,Simon&al}. Our main result
is that, in the limit of a large window width $\Dt$, the entropy increment produced 
during time $\Dt$ approaches a constant
(or, at least, a very slowly varying function of $\Dt$).
Dividing this constant by $\Dt$ and
taking the limit $\Dt \to \infty$, we find that the entropy flux is zero.

We find this result interesting in two respects. First, it supports the interpretation
of the entropy flow as a flow of information: the flow stops once all
the information about the initial state of the paramp has been released. 
Second, it provides an example of how a continuous signal can be mapped into a sequence
of multimode quantum states (the modes here being the components of the windowed
cosine transform). This connects the more traditional methods of quantum circuit theory,
such as the input-output formalism, to quantum information theory (where a signal is 
a stream of density matrices). In the present case (vacuum input),
the sequential output states are approximately multimode squeezed vacua, 
with $1/ \Dt$ corrections representing 
entanglement across the sequence.

Next, we connect the vanishing of the entropy flux from the paramp to development 
of the off-diagonal coherences $\la B_k B_{k'} \ra$ of the output operators. 
These coherences do not participate
in the computation of the energy or photon number fluxes, but play an essential role
in the computation of the entropy flux. 
We argue that they represent transfer of entanglement from the paramp mode
to the output. 

Recall that off-diagonal coherences are absent from 
Hawking's calculation \cite{Hawking,Hawking:1976} of radiation from a black hole.
Many authors had argued that small corrections to Hawking's result
can accumulate and produce a result for the entropy of the radiation 
that decreases at large times; for a recent review of the argument, 
see Ref.~\cite{Almheiri&al}.
The present study motivates us to look for a version of entanglement transfer that
may be relevant in that case. The one we find looks very  
similar to the entanglement swapping protocol
of quantum optics \cite{Zukowski&al}.

Finally, we propose to use 
measurements of the off-diagonal coherences as stand-ins for measurements of the information
flow, with the purpose of finding out experimentally if (as suggested by work \cite{Page}
on Hawking radiation ) high-entropy subsystems release
information to radiation much more slowly than low-entropy ones do.

The paper is organized as follows. In Sec.~\ref{sec:deg}, we review properties of the
degenerate paramp relevant to the present work. We assume that the paramp is
operated close to resonance but not necessarily directly at it, namely, that  
the frequency detuning $\dome$ satisfies
$|\dome| < f$, where $f$ is the amplitude of the pump. This translates into a generally 
nonzero value of the angle $\phi$ defined by 
$\sin 2\phi = \dome / f$ and is done so as to make sure that the conclusions we reach are
not tied specifically to the resonance condition $\dome = 0$. In Sec.~\ref{sec:cor} we
define correlation functions required for computations of the entropy. The computations 
themselves are presented in Sec.~\ref{sec:ent1} (for the paramp) and Sec.~\ref{sec:ent2} 
(for the output). Sec.~\ref{sec:ent2} ends with a comparison between 
the computations of the output entropy flux and of the fluxes of 
the energy and number of quanta. Mechanisms of entanglement transfer
are discussed in Sec.~\ref{sec:mech}. Sec.~\ref{sec:concl} is a conclusion.

\section{Degenerate paramp} \label{sec:deg}

A driven degenerate paramp is described by the Hamiltonian \cite{Collett&Gardiner}
\begin{multline}
H = \ome_s a^\dagger a 
+ \frac{i}{2} f \left[ e^{- i \ome_p t}  (a^\dagger)^2 - e^{i \ome_p t} a^2 \right] \\
{} 
+ \sum_\nu \left[ \eps_\nu b_\nu^\dagger b_\nu + i g_\nu ( b_\nu^\dagger a - a^\dagger b_\nu) \right] .
\label{ham}
\end{multline}
This includes the paramp itself, represented by operator $a$, with signal frequency $\ome_s$,
driven by a classical pump with
amplitude $f$ and frequency $\ome_p$ and coupled to a bath of oscillators with frequencies
$\eps_\nu$. The coupling constants $g_\nu$ are assumed real, and the drive 
amplitude $f$ real and
positive. All operators obey the canonical Bose commutation relations. We set $\hbar = 1$.

The Markov approximation corresponds to the replacement
\be
\sum_\nu g_\nu^2 e^{-i \eps_\nu (t-t')} \to \Gam \delta(t - t') \, ,
\label{mark}
\ee
where $\Gam$ is a positive constant. In the frequency domain, this corresponds to a constant
(white-noise) power spectrum for the sum on the left-hand side. Eq.~(\ref{mark}) is the form 
of the approximation we use in what
follows: there will be no need to impose constancy of $g_\nu$ and of the density of states 
separately.

We employ the input-output formalism described in Ref.~\cite{Gardiner&Collett}, 
except that we do not assume from the outset 
that the bath spectrum is continuous. The input operators are defined as
\be
\bin(t) = \frac{1}{\sqrt{\Gam}} \sum_\nu g_\nu e^{-i \eps_\nu t} b_\nu(0) \, ,
\label{bin}
\ee
where $t > 0$, and
$b_\nu(0)$ is the initial datum for the Heisenberg operator $b_\nu(t)$ of the bath. 
Note that the single
$\bin$ subsumes the effect of many $b_\nu$. As a consequence of (\ref{mark}), 
the input operators satisfy the commutation relation
\be
[ \bin(t) , \bin^\dagger(t') ] = \delta(t-t') \, .
\label{bcom}
\ee
Note that, apart from (\ref{mark}), 
this result relies only on the equal-time commutators of $b_\nu$ (namely, those at
$t=0$).

Next, following the steps outlined in Ref.~\cite{Gardiner&Collett}, 
we obtain the Heisenberg (quantum Langevin) equation of motion for the paramp 
operator $a$:
\[
\dot{a} = -i \ome_s a - \half \Gam a + f e^{-i \ome_p t} a^\dagger - \sqrt{\Gam} \bin \, .
\]
It is convenient to switch to the rotating frame by defining the tilded operators $\ta$ via
$a(t) = e^{-i\ome_p t / 2} \ta(t)$ and similarly for $\bin$. The equation becomes
\be
\dot{\ta} = - i (\dome) \ta - \half \Gam \ta + f \ta^\dagger - \sqrt{\Gam} \tbin \, ,
\label{eqa}
\ee
where $\dome = \ome_s - \ome_p / 2$.

If the paramp is sufficiently close to resonance, i.e., $|\dome| < f$, 
we can search for a solution to the corresponding homogeneous equation
in the form $\ta(t) = c_1 e^{\lam_1 t} + c_2 e^{\lam_2 t}$, where 
\be
\lam_1 = - \half \Gam + f' \, , \hspace{3em}
\lam_2 = - \half \Gam - f' \, , 
\label{lam}
\ee
and 
\be
f' = [f^2 - (\dome)^2]^{1/2} \, .
\label{f'}
\ee
We will assume that the paramp is operated in this regime and 
below the instability threshold, $f' < \Gam / 2$. Then, both $\lam_1$ and $\lam_2$
are negative.

The solution to Eq.~(\ref{eqa}) can be expressed in terms of the Hermitian quadratures
\ba
X_1(t) & = &  \cN^{1/2} [  e^{-i\phi} \ta(t) +  e^{i\phi} \ta^\dagger(t) ]  \, , \label{X1} \\
X_2(t) & = &  i \cN^{1/2} [  e^{-i\phi} \ta^\dagger(t) - e^{i\phi} \ta(t) ] \, , \label{X2}
\ea
where $\phi$ is the angle in the range $|\phi| < \pi / 4$ defined by
\be
f' + i (\dome) = f e^{2 i \phi} \, ,
\label{phi}
\ee
and
\be
\cN = (\cos 2\phi)^{-1} = f / f' \, .
\label{cN}
\ee
The on-resonance case $\dome = 0$ corresponds to $\phi = 0$. 

The solution in question is
\be
X_\a(t) = X_\a(0) e^{\lam_\a t} - \sqrt{\Gam} \int_0^t ds Y_\a(s) e^{\lam_\a (t - s)} \, ,
\label{Xt}
\ee
where $\a = 1,2$, and $Y_\a$ are the quadratures of the input 
defined similarly to (\ref{X1}), (\ref{X2}):
\ba
Y_1(t) & = & \cN^{1/2} [ e^{-i\phi} \tbin(t) + e^{i\phi} \tbin^\dagger(t) ] \, , \label{Y1} \\ 
Y_2(t) & = & i \cN^{1/2} [ e^{-i\phi} \tbin^\dagger(t) - e^{i\phi} \tbin(t) ] \, . \label{Y2}
\ea
The commutation relations of $Y_\a$ follow from (\ref{bcom}): 
\be
[Y_1(t), Y_2(t')] = 2 i \delta(t-t') \, ,
\label{Ycom}
\ee
and those of $X_\a$ from (\ref{Xt}). In particular, the equal-time commutator 
is $[X_1(t), X_2(t)]  = 2 i$.

The output of the paramp is described by the operators
\be
\bout(t) = \frac{1}{\sqrt{\Gam}} \sum_\nu g_\nu e^{-i \eps_\nu (t - \tp)} b_\nu(\tp) \, ,
\label{bout_def}
\ee
where $\tp$ is in the future of $t$; they satisfy the relation \cite{Gardiner&Collett}
\be
\bout(t) = \sqrt{\Gam} a(t) + \bin(t) 
\label{bout}
\ee
and the same commutation relation as $\bin$: 
\be
[\bout(t), \bout^\dagger(t')] = \delta(t - t') \, .
\label{out_com}
\ee
The rotating-frame version of these
operators is $\tbout(t) = e^{i \ome_p t / 2} \bout(t)$.
The customary interpretation of (\ref{bout}) is that it describes the effect of
a ``collision'' of the bath with the paramp at time $t$, with $\bin$ ($\bout$) 
representing the state of the bath before (after) the collision (see, for example, 
Ref.~\cite{Clerk&al}).

\section{Correlation functions} \label{sec:cor}

In general, the initial state of the bath may have some useful input to be amplified, in addition
to the inevitable noise. Here, we present a computation of the entropies of various subsystems for
the case when the initial state is pure zero-temperature noise, that is, the vacuum 
of the operators $b_\nu(0)$.  
In the Markov approximation (\ref{mark}), this leads to
\be
\la \bin(t) \bin^\dagger(t') \ra = \delta(t - t') \, ,
\label{vac}
\ee
with the expectation values of the other bilinears of $\bin$, $\bin^\dagger$ vanishing. 

Computation of the entropies uses correlation functions of
the quadratures defined in the preceding section. The (unsymmetrized) two-point
correlation function of
the input quadratures $Y_\a$ follows directly from (\ref{vac}):
\be
\cor{Y}_{\a\b}(t, t') = \la Y_\a(t) Y_\b(t') \ra 
= \delta(t - t') M_{\a\b} \, ,
\label{corY}
\ee
where $M_{\a\b}$ are the matrix elements of
\be
\hat{M} = \cN \left( \begin{array}{cc} 1 & i e^{-2i\phi} \\ -i e^{2i\phi} & 1 \end{array}
\right) .
\label{hM}
\ee
Next, using the solution (\ref{Xt}), we compute the correlation matrix of $X_\a$,
defined similarly to the above, i.e., as $\cor{X}_{\a\b}(t, t') = \la X_\a(t) X_\b(t') \ra$:
\begin{multline}
\cor{X}_{\a\b}(t, t') = e^{\lam_\a t + \lam_\b t'} \bigg\{ \cor{X}_{\a\b}(0,0)  \\
{} + \frac{\Gam M_{\a\b}}{\lam_\a + \lam_\b} \left[ 1 - e^{- (\lam_\a + \lam_\b) t_<} \right] 
\bigg\} ,
\label{corX}
\end{multline}
where $t_<$ is the smaller of $t$ and $t'$. Finally, to compute the entropy of the output,
we will need the correlation matrix 
$\cor{Z}_{\a\b}(t, t') = \la Z_\a(t) Z_\b(t') \ra$ of the output quadratures
\be
Z_\a(t) = \sqrt{\Gam} X_\a(t) + Y_\a(t) \, ,
\label{Zquad}
\ee
which are related to the output operators $\tbout$, $\tbout^\dag$ by formulas similar to
(\ref{Y1}), (\ref{Y2}). That correlation matrix is
\be
\cor{Z}_{\a\b}(t, t') = \Gam \cor{X}_{\a\b}(t, t') - \Gam h_{\a\b}(t,t') M_{\a\b}
+ \cor{Y}_{\a\b}(t, t') \, ,
\label{corZ}
\ee
where we have defined the function
\be
h_{\a\b}(t,t') = \left\{ \begin{array}{cc} e^{\lam_\a (t - t')} \, , & t > t' \\
e^{\lam_\b (t' - t)} \, , & t < t' \, . \end{array} \right.
\label{hfunc}
\ee
Note that the same function occurs also in the correlator of $X_\a$, Eq.~(\ref{corX}), 
where in the term containing $t_<$
\be
\exp [ \lam_\a t + \lam_\b t' - (\lam_\a + \lam_\b) t_< ] = h_{\a\b}(t,t') \, ,
\label{hfunc2}
\ee
and that it is this term that determines the properties of the correlator 
at large times.

\section{Entanglement entropy of the paramp} \label{sec:ent1}

We begin with the entropy of the paramp itself. For the computation at early times,
we will assume that the initial state of the paramp is Gaussian with zero mean for 
either quadrature. Such a state is completely characterized by the correlation matrix
$\cor{X}_{\a\b}(0,0)$ and
will remain Gaussian with zero mean under the evolution described by
(\ref{eqa}) (for a review of Gaussianity-preserving quantum
Langevin dynamics, see Ref.~\cite{Genoni&al}).
At late times, we will not need the assumption of Gaussianity of the initial state, 
as the initial data will be completely replaced by noise from
the input line.

The calculation now is entirely standard. It is based on the fact that the entanglement
entropy is immune to canonical transformations. A convenient object to apply such
a transformation to is the covariance matrix $\hC(t)$, obtained from the correlation
matrix (\ref{corX}) by setting $t'=t$ and symmetrizing:
\be
C_{\a\b}(t) = \half [ \cor{X}_{\a\b}(t, t) +  \cor{X}_{\b\a}(t, t) ] \, .
\label{cov}
\ee
We find
\be
C_{\a\b}(t) =   C_{\a\b}(0) e^{(\lam_\a + \lam_\b) t}
+ \frac{\Gam  N_{\a\b}}{\lam_\a + \lam_\b} \left[ e^{(\lam_\a + \lam_\b) t} - 1 \right]  \, ,
\label{cov2}
\ee
where $N_{\a\b}$ are the matrix elements of 
\be
\hat{N} = \cN \left( \begin{array}{cc} 1 & \sin 2\phi \\ \sin 2\phi & 1 \end{array}
\right) , \hspace{3em} \cN = \frac{1}{\cos 2\phi} \, .
\label{hN}
\ee
The $2\times 2$ symmetric matrix $\hC$ can be brought to a certain normal 
form---a diagonal matrix with equal entries on the diagonal---by a combination 
of an orthogonal
rotation and a rescaling $X_1 \to \mu X_1$, $X_2 \to \mu^{-1} X_2$ 
(a Bogoliubov transformation). This is a limiting case of the transition to the
Williamson normal form \cite{Williamson} in the multimode case \cite{Simon&al}.

Since neither of the transformations mentioned above changes the determinant, 
the diagonal element $\gam$ of the normal form can be computed from
\be
\gam^2(t) = \det \hC(t) \, .
\label{gam}
\ee
The normal form can be thought of as representing a thermal state of the fictitious
instantaneous Hamiltonian ${\cal H}(t) = p^2(t) + q^2(t)$, where $q(t)$ and $p(t)$ 
are the quadratures obtained as a result of the transformations.
The mean level occupancy in this state is given 
(in our normalization of the quadratures) by
\be
\nth(t) = \half [\gam(t) - 1 ] \, .
\label{nth}
\ee
The entanglement entropy of the paramp at time $t$
can then be found from the ideal-gas von Neumann formula 
\be
S_{par}  = (\nth + 1) \ln (\nth + 1) - \nth \ln \nth \, .
\label{EE}
\ee

Let us consider some asymptotics. At small times, to the linear order in $t$, 
\begin{multline}
\det \hC(t) \approx \det \hC(0) (1 - 2 \Gam t) \\
{} + \Gam \cN t \left[ C_{11}(0) + C_{22}(0) - 2 C_{12}(0) \sin 2\phi \right] \, ,
\label{det}
\end{multline}
where we have used the relation $\lam_1 + \lam_2 = -\Gam$. Unless the linear in $t$
term in this expression vanishes, Eq.~(\ref{det}) determines the leading term in 
the entropy growth at small $t$. 

For example, suppose the initial state is a squeezed vacuum of $\ta(0) = a(0)$, with
squeezing by an amount $r$ applied in the direction of the real part of $a(0)$. 
The covariance matrix of the quadratures
(\ref{X1}), (\ref{X2}) in this state is
\begin{widetext}
\be
\hC(0) = \cN \left( \begin{array}{cc} 
\cosh 2r + \sinh (2r) \cos 2\phi & \cosh (2r) \sin 2 \phi \\
\cosh (2r) \sin 2 \phi & \cosh 2r - \sinh (2r) \cos 2\phi 
\end{array} \right) .
\label{Csq}
\ee
\end{widetext}
As for any pure state, $\det \hC(0) = 1$. Substituting the matrix elements of (\ref{Csq})
in (\ref{det}), we obtain
\be
\det \hC(t) \approx 1 + 4 \Gam t \sinh^2 r  \, .
\label{stream}
\ee
To logarithmic accuracy, the leading contribution to the entropy at small $t$
then comes from the second term in (\ref{EE}) and equals
\be
S_{par}(t) \approx - (\Gam t \sinh^2 r) \ln t  \, .
\label{log}
\ee
Note the characteristic $t \ln t$ behavior \cite{HBosons}. 
Note also that (\ref{log})
depends only on $\Gam$ but not on $\lam_{1,2}$ separately. This suggests that at early times 
the presence of the drive is immaterial, and the growth of the entropy can
be attributed to free streaming of quanta out of the paramp. This makes it 
analogous to the growth of the entanglement entropy in a (comoving) region of an
expanding universe after a transition from inflation to 
a slower expansion---or, as a limiting case, to a static universe \cite{fheat}. 
In either case, one can think of the squeezed initial state as containing ``latent''
quanta, which are released by the subsequent evolution. This interpretation
of (\ref{log}) 
is supported by the presence of the $\sinh^2 r$ factor, which coincides with the
average number of $a$ quanta in the initial state.

Next, we consider the limit of large times, $t \gg 1/|\lam_1|$.
At a large enough $t$, the dependence
on the initial state is lost, and $\hC(t)$ approaches a constant matrix, 
with the matrix elements given by
\be
C_{\a\b} (t \to \infty) = - \frac{\Gam N_{\a\b}}{\lam_\a + \lam_\b} \, .
\label{Cinf}
\ee
This means that the final-state entropy is also a constant.
The determinant of (\ref{Cinf}) is
\be
\det \hC(t \to \infty) = \cN^2 \left( \frac{\Gam^2}{4 \lam_1 \lam_2} - \sin^2 2 \phi \right)
= 1 + \frac{f^2}{\lam_1\lam_2} \, ,
\label{det2}
\ee
where $\lam_1 \lam_2 = \frac{1}{4} \Gam^2 - f^2 + (\dome)^2$.
We see that, unlike the entropy growth at early times, 
having a nonzero entropy at $t\to \infty$ relies entirely on the presence of the drive,
$f > 0$. Note also that, by bringing the paramp close to the instability threshold, 
i.e., $\lam_1$ close to zero, one can make the final-state
entropy arbitrarily large.

It is of interest to trace the transition from the early to late-time asymptotic 
behavior for the initial state (\ref{Csq}) with $\sinh^2 r \gg 1$. This is the case 
when we may expect the initial free streaming of quanta to rapidly
produce a large entropy. We will consider that transition here for positive $r$ and
a paramp operated precisely at resonance ($\dome = 0$). 
Then, $\phi = 0$ (so that the initial
squeezing is in the direction of the $X_1$ quadrature), and the covariance
matrix constructed from (\ref{cov2}) is diagonal:
\begin{widetext}
\be
\hC(t) = \left( 
\begin{array}{cc} e^{2r + 2 \lam_1 t} + \frac{\Gam}{2\lam_1} (e^{2\lam_1 t} - 1) & 0 \\
0 & e^{-2r + 2 \lam_2 t} + \frac{\Gam}{2\lam_2} (e^{2\lam_2 t} - 1) \end{array} \right) .
\label{cov3}
\ee
\end{widetext}
Neglecting $e^{-2r}$ in comparison with unity, we obtain
\be
\det \hC(t) \approx  \frac{\Gam^2}{4\lam_1 \lam_2} \left[ 1 - e^{2\lam_2 t} +
(1 - x) \left( e^{-2\Gam t} - e^{2 \lam_1 t} \right)
\right] ,
\label{det3}
\ee
where
\be
x \equiv \frac{2 |\lam_1| e^{2r}}{\Gam} 
\label{xpar}
\ee
is the parameter that contains all the information about the initial state (that is, 
about the value of $r$). Because of cancellations at $t=0$, (\ref{det3}) 
is not a good approximation
at very small times, but quickly becomes one after the second term in (\ref{stream}) 
overtakes the first. In addition, at not too large times, $t \ll e^{2r} / \Gam$,
we can replace $e^{2\lam_2 t}$ in (\ref{det3}) with $e^{-2 \Gam t}$, since the difference
\be
e^{2\lam_2 t} - e^{-2 \Gam t} = - e^{-2 \Gam t} \left( 2 \lam_1 t + \dots \right)
\label{diff}
\ee
is much smaller than the $x e^{-2\Gam t}$ term already present in (\ref{det3}). 
The result is
\be
\det \hC(t) \approx  \frac{\Gam^2}{4\lam_1 \lam_2} \left[ 1 - x e^{-2\Gam t} +
(x - 1) e^{2 \lam_1 t} \right] ,
\label{det4}
\ee
which can be easily analyzed. In particular, for $x > 1$, this expression (and hence also
the entropy) has a maximum at
\be
t_{\max} = \frac{1}{2|\lam_2|} \ln \frac{\Gam x}{|\lam_1| (x - 1)} \, .
\label{tmax}
\ee
Note that, for small $|\lam_1|$,
the rate constant at which (\ref{det4}) decays from the maximum at $t = t_{\rm max}$
to the asymptotic value (\ref{det2}), and the paramp ``forgets'' the initial state, 
is set by $|\lam_1|$ and is exponentially small in the final-state entropy. 

\begin{figure}
\begin{center}
\includegraphics[width=3.25in]{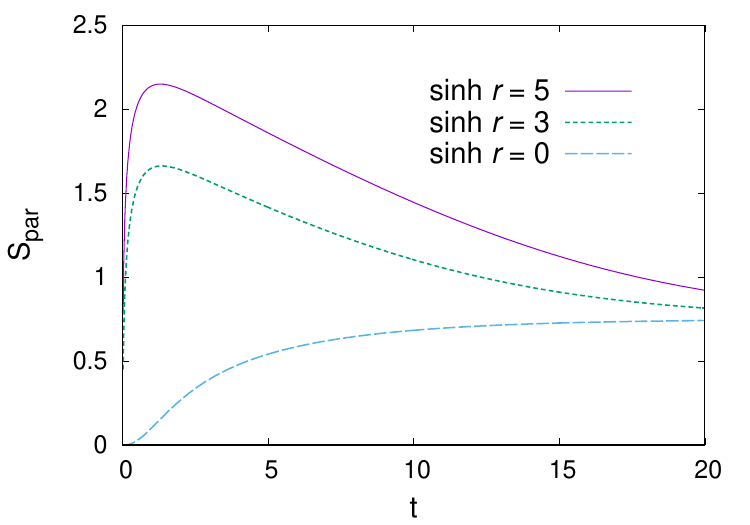}
\end{center}
\caption{\small Entanglement entropy, as a function of time, of the 
paramp with $\Gam = 1$, $f= 0.4$ operated at resonance ($\dome = 0$) and starting
out in a squeezed vacuum described by the covariance matrix (\ref{Csq}) with $\phi = 0$ 
and various values of $r$.}
\label{fig:Spar}
\end{figure}

In Fig.~\ref{fig:Spar}, we plot the entropy (\ref{EE}) as function of time for a paramp 
operated precisely
at resonance ($\dome = 0$) and fairly close to the threshold, for a family of squeezed
initial states described by (\ref{Csq}) with different values of $r$. We see that, for 
larger $r$, there are indeed two distinct stages: the rapid early growth, followed
by a slower later stage. The overall time dependence of $S_{par}$ in this case is 
reminiscent of that found in Ref.~\cite{HBosons} for the entanglement entropy of 
a subsystem during thermalization of an isolated system.

\section{Entropy of the output} \label{sec:ent2}

\subsection{Discretization}

We now turn to computation of the entropy associated with the output quadratures 
(\ref{Zquad}). The presence of the delta function in (\ref{corY}) and, as a consequence,
in (\ref{corZ}) means that we cannot 
apply the standard methods directly. We need to discretize the output signal first.
We do that by defining windowed cosine transforms of the output quadratures (\ref{Zquad}):
\be
Z_\a^{(jk)} = \int_{t_j}^{t_{j+1}} Z_\a(t) u_{jk}(t) dt \, ,
\label{Zjk}
\ee
where $t_j = j \Dt$ are equally spaced moments of time, labeled by an integer 
$j \geq 0$ 
and separated by a window width $\Dt$, and
\be
u_{jk}(t) = \frac{\eta_k}{\sqrt{\Dt}} \cos \left[ \ome_k (t - t_j) \right]
\label{ujk}
\ee
with
\be
\ome_k = \frac{\pi k}{\Dt} \, , \hspace{3em} k = 0, 1, \dots ,
\label{ome}
\ee
and 
\be
\eta_k = \left\{ \begin{array}{rl} 1, & k = 0 \\ \sqrt{2}, &  k > 0 \, . \end{array} \right.
\label{eta}
\ee
Discretized versions $B_{jk}$ of the operators $\tbout(t)$
are defined similarly. They are related to the operators (\ref{Zjk}) by
\ba
B_{jk} & = & \half \cN^{1/2} \left[ Z_1^{(jk)} e^{-i\phi} + i Z_2^{(jk)} e^{i\phi} \right] \, ,
\label{Bjk} \\
B_{jk}^\dag & = & 
\half \cN^{1/2} \left[ Z_1^{(jk)} e^{i\phi} - i Z_2^{(jk)} e^{-i\phi} \right]
\label{Bjk_dag}
\ea
and satisfy the canonical commutation relations. 
These operators are quantum analogs of Gabor's ``elementary signals'' \cite{Gabor}
of the classical information theory.

A windowed Fourier transform has been used in Ref.~\cite{modes_of_univ} to study
squeezing in the output of laser and in Ref.~\cite{Clerk&al} 
as a means of computing the number flux of photons on a transmission line. 
Here, we use it to compute the entropy flux in the output of 
a paramp (we will discuss the number and energy fluxes as well).

We now consider the quadratures corresponding to a given time window, i.e., $Z_\a^{(jk)}$ 
all having the same $j$ but various $k$. To keep the number of degrees of
freedom finite, we assume that $k$ goes over the range
\be
k = 0, \dots , \kmax 
\label{kmax}
\ee
for some large $\kmax$. In the end, we will have to see if the result for the entropy has 
a limit at $\kmax \to \infty$.

The restriction to a single time window means that we lose information about the
quadratures at other times. We can then expect that there will be a ``boundary'' 
contribution to the entropy, coming from vicinity of the window endpoints. 
The question we wish to ask
is if there is also a ``bulk'' contribution, coming from the entire length of the
window and perhaps such that at large $\Dt$ the entropy, $\DSout$,
becomes proportional to $\Dt$ 
(similarly to the volume law for the entropy of an excited state in the case of a spatial
partition). If that turns out to be the case, we can attribute to the ratio $\DSout/ \Dt$
the significance of the entropy flux (similarly to how the corresponding quantity for 
a spatial partition is given the significance of the entropy density).
Since correlations in (\ref{corZ}) extend over time scales 
$t \sim 1 / |\lam_1|$ ($\lam_1$ being the less negative of the two $\lam_\a$), to be able
to tell a bulk contribution from a boundary one, we must use a window width exceeding
that timescale.

\subsection{Computation of the entropy flow}

We now focus on the limit of large window widths, $\Dt |\lam_1| \gg 1$, which means
that we will neglect
quantities that are exponentially small in this parameter. This limit also allows us 
to neglect, to the same accuracy, the initial transient in (\ref{corX}). Since
that amounts to the assumption that the initial state of the paramp is forgotten, 
we may as well assume the initial state to be Gaussian with zero mean for all quadratures. 
The combined system $\{a, b_\nu \}$ will then undergo Gaussian evolution, described
by the Hamiltonian (\ref{ham}) and resulting in a Gaussian state of the final-state 
operators $b_\nu(t_+)$ that appear in (\ref{bout_def}) (for a review of Gaussian quantum
dynamics, see Ref.~\cite{Genoni&al}). This means
that correlators of multiple instances of $b_\nu(t_+)$ and $b_{\nu'}^\dag(t_+)$
can be expressed through the pairwise correlators by the Wick-Gaudin theorem \cite{Gaudin}.
The same is true then for the operators $\tbout(t)$, and also for $B_{jk}$. So,  the state of
the operators $B_{jk}$ is Gaussian (with zero mean), and for computation of the
entropy of that state we can apply the method based on symplectic diagonalization 
\cite{Williamson,Simon&al} of its covariance matrix.

With the initial transient neglected, the correlation matrix (\ref{corX}) of $X_\a$ 
becomes
\be
\cor{X}_{\a\b}(t, t') \approx - \frac{\Gam M_{\a\b}}{\lam_\a + \lam_\b} h_{\a\b}(t, t') \, .
\label{corX2}
\ee
Substituting this in the expression (\ref{corZ}) for the correlation matrix of $Z_\a$, 
we obtain
\be
\cor{Z}_{\a\b}(t, t') \approx \Gam L_{\a\b} h_{\a\b}(t, t')  + M_{\a\b} \delta(t-t') \, ,
\label{corZ2}
\ee
where $L_{\a\b}$ are the elements of the diagonal matrix
\be
\hL = \left( \begin{array}{cc} 
 - \frac{f}{\lam_1}  & 0 \\ 0 & \frac{f}{\lam_2} 
\end{array} \right) ,
\label{hL}
\ee
and $M_{\a\b}$ are the matrix elements of (\ref{hM}). Applying the cosine transform
(\ref{Zjk}) and symmetrizing, we obtain the covariance matrix of the quadratures 
$Z_{\a k} \equiv Z_\a^{(jk)}$ as
\be
\hC^{(out)}_{kk'} \approx 2 \Gam f \left( \begin{array}{cc} F_{kk'}(1) & 0 \\ 0 & - F_{kk'}(2) 
\end{array} \right) + \hat{N} \delta_{kk'} \, ,
\label{Cout}
\ee
where the approximation sign indicates that terms suppressed exponentially 
in $|\lam_\a| \D t$ have been neglected,
\be
F_{kk'}(\a) = \frac{\delta_{kk'}}{\lam_\a^2 + \ome_k^2} - 
\frac{|\lam_\a| \cos^2 \! \left[ \frac{\pi}{2} (k + k') \right]}
{\Dt (\lam_\a^2 + \ome_k^2)(\lam_\a^2 + \ome_{k'}^2)} \eta_k \eta_{k'} ,
\label{F}
\ee
$\ome_k$ and $\eta_k$ are given by (\ref{ome}), (\ref{eta}), 
and $\hat{N}$ is the matrix (\ref{hN}). The cosine squared in (\ref{F}) 
is nonzero only when $k+k'$ is even. 
In other words, the covariance matrix factorizes into two blocks:
one for the even $k$, and the other for the odd.

The second term in (\ref{F}) vanishes as $1/ \Dt$ in the limit $\Dt \to \infty$, 
which may be taken to identify it as a contribution from vicinity of the window endpoints.  
Still, the effect of this
term needs to be considered carefully, as there is a large number of such terms 
(occupying, more or less, a square with a side length of order
$|\lam_\a|$ in the $\ome_k$, $\ome_{k'}$ plane). 

Note also the nontrivial values of the off-diagonal coherences 
of the operators $B_k \equiv B_{jk}$,
\be
\la B_k B_{k'} \ra \approx  \half  \Gam f \cN
\left[ e^{-2i\phi} F_{kk'}(1) + e^{2i\phi} F_{kk'}(2) \right] ,
\label{Boff}
\ee
which can be found on the basis of (\ref{Cout}).

The symplectic eigenvalues $\gam_\ell$ of (\ref{Cout}), with both $k$ and $k'$ 
restricted to the range (\ref{kmax}), can now be used
to compute the entropy by the ideal-gas formula
\be
\D S_{out} = \sum_\ell \left[ (n_\ell + 1) \ln (n_\ell + 1) - n_\ell \ln n_\ell \right] \, ,
\label{DS}
\ee
where $n_\ell = \half (\gam_\ell - 1)$ [compare to (\ref{nth})]. 
To see why one may hope to obtain in this way a 
quantity with the
significance of the entropy flux, i.e., the amount of entropy output per unit time,
suppose for the moment that the first term in (\ref{F}) is all there is. 
With the second term in (\ref{F}) thus neglected, the matrix (\ref{Cout}) becomes diagonal
in $k$, $k'$, i.e., factorizes into $2\times 2$ blocks, each having the form
\be
\hC(\ome_k) = \cN \left( \begin{array}{cc} 
1 + \frac{2 \Gam f'}{\lam_1^2 + \ome_k^2} & \sin 2\phi \\
\sin 2\phi & 1 - \frac{2 \Gam f'}{\lam_2^2 + \ome_k^2} \end{array} \right) .
\label{Ck}
\ee
The entanglement entropy $\DSout$ of the output 
can then be written as the sum of the entropies of these blocks:
\be
\D S_{out} = \sum_{k=0}^{\kmax} S(\ome_k) 
\to \frac{\D t}{\pi} \int_0^\infty d\ome S(\ome) \, ,
\label{DS2}
\ee
where in the last expression we have taken $\kmax$ to infinity and
replaced the sum with an integral. 
If the integral is finite, it defines a finite entropy flux $\D S_{out} / \Dt$. 

The entropy of each $2\times 2$ block (\ref{Ck}) can be computed using the principle
we used earlier for the paramp itself. That is, the matrix (\ref{Ck}) can be brought 
to the Williamson normal form  
by a combination of an orthogonal rotation and a Bogoliubov transformation. 
Since both of these preserve the determinant, the symplectic eigenvalue $\gam_k$
can be found from $\gam_k^2 = \det \hC(\ome_k)$.
A direct computation shows that the determinant of (\ref{Ck})
is unity, meaning that the state is pure (a squeezed vacuum). 
In other words, in the approximation where the second term 
in (\ref{F}) is neglected, there is no entropy flow out of the paramp.
This is a surprising result, given that, as we will see shortly,
the number and energy flows (in the same approximation) are both nonvanishing.

\begin{figure}
\begin{center}
\includegraphics[width=3.25in]{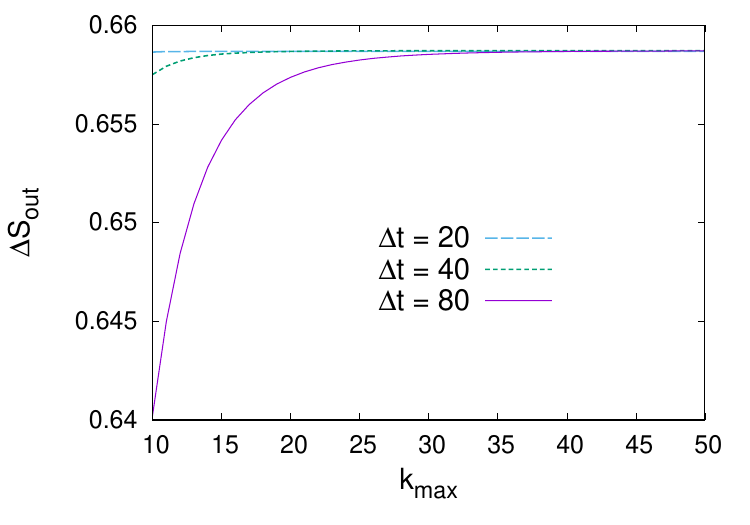}
\end{center}
\caption{\small Entanglement entropy of the output signal in a given time 
window, computed numerically by symplectic diagonalization
of the covariance matrix (\ref{Cout}) for three different window widths, as a function
of the number of included harmonics.
The plots illustrate both the convergence of the
results in the limit of large $\kmax$ and the independence of the value
of $\DSout$ in that limit from $\Dt$ at large $\Dt$. 
Here, the paramp with $\Gam = 1$ and $f = 0.3$ is operated off-resonance, at $f' = 0.2$.}
\label{fig:DSout}
\end{figure}

We now return to the complete expression (\ref{F}). A numerical computation of the
symplectic eigenvalues $\gam_\ell$ of (\ref{Cout}) brings a further surprise: for each
of the two matrices contained in (\ref{Cout}) (one with even $k$, the other with odd),
only one eigenvalue is significantly different from unity;
for the rest, the differences 
$\gam_\ell - 1$ are of order $10^{-14}$ or less, 
which we judge to be zero. As a result,
the sum in (\ref{DS}) contains only two terms. The two nontrivial $\gam_\ell$ are both 
larger than unity, as required \cite{Simon&al} in order for $\hC^{(out)}$ to be
consistent with the uncertainty principle. They differ somewhat from 
each other for smaller $\kmax$ but rapidly approach 
a common limit as $\kmax$ is increased. Thus, $\DSout$ appears to have 
a good limit at $\kmax \to \infty$. Moreover, this limit becomes virtually independent
of $\Dt$ at large $\Dt$. These properties of the numerical results are illustrated in
Fig.~\ref{fig:DSout}. 

Because $\overline{\D S}_{out}$, 
by which we now denote the limit of $\DSout$ at large $\kmax$, 
approaches a constant 
(or, in any case, a very slow varying function of $\Dt$) at large $\Dt$,  
\be
\lim_{\D t \to\infty} \frac{\overline{\D S}_{out}}{\D t} = 0 \, ,
\label{lim}
\ee
meaning that the entropy flux (the flow rate) is zero.

The independence of $\overline{\D S}_{out}$ from $\Dt$ at large $\Dt$ is consistent with 
it being accumulated entirely near the window endpoints. Indeed,
such ``boundary'' contributions to the entropy have been anticipated in our initial 
discussion of the transformation (\ref{Zjk}). We also note the following curious coincidence: 
numerically, the value of $\overline{\D S}_{out}$ is, to a good accuracy, 
twice the asymptotic entropy of the paramp, 
$S_{par}(t\to\infty)$, computed in Sec.~\ref{sec:ent1}. 

\subsection{Comparison with the flows of other quantities}

Unlike the computation of the entropy flux,
computation of the flux of the number of particles requires access only to the diagonal 
coherences 
\be
\la B^\dagger_k B_{k'} \ra \approx \half  \kap \left[ F_{kk'}(1) - F_{kk'}(2) \right] ,
\label{Bdiag}
\ee
where $\kap = \Gam f \cN$, and $F_{kk'}$ is given by (\ref{F}).
As before, $B_k \equiv B_{jk}$, and
the approximation sign indicates the large-$\Dt$ limit. We first compute the 
particle number increment
\be
\D N_{out} = \sum_{k=0}^\infty \la B^\dagger_k B_k \ra \, .
\label{DN}
\ee
Here, the contribution of the second term in (\ref{F}) is clearly subleading at large 
$\D t |\lam_\a|$. Neglecting that term and
replacing the sum with an integral, we find the number flux to be
\be
\frac{\D N_{out}}{\Dt} = \frac{\Gam f^2}{2 \lam_1 \lam_2} \, .
\label{DN2}
\ee
This expression can in fact be obtained without any windowing, simply
as the large-$t$ limit of the equal-time correlator
$\la \tbout^\dagger(t) \tbout(t) \ra$. The latter is found by setting $t=t'$ in 
\be
\la \tbout^\dagger(t) \tbout(t') \ra \approx \frac{1}{4} \kap \left[ 
\frac{1}{|\lam_1|} h_{11}(t,t') - \frac{1}{|\lam_2|} h_{22}(t,t') \right] , 
\label{bdiag}
\ee
which follows directly from (\ref{corZ2}) and 
\be
\tbout(t) = \half \cN^{1/2} \left[ Z_1(t) e^{-i\phi} + i Z_2(t) e^{i\phi} \right] .
\label{tbout2}
\ee

Similarly, the energy flux can be computed from the expectation value of 
the operator
\be
P_{out} = 
\frac{i}{2} \left( \bout^\dag \part_t \bout - \part_t \bout^\dag \bout \right) \, .
\label{Pout}
\ee
This operator generates time translations of
$\bout$ in the sense that, for any smooth test function $f(t)$ 
that has a vanishing derivative at $t=\tau$,
\be
\left[ \int P_{out}(t) f(t) dt, \bout(\tau) \right] = - i f(\tau) \part_\tau \bout(\tau) \, .
\label{gen}
\ee
Choosing $f(t)$ to approach the indicator function of the interval $(t_j, t_{j+1})$,
we can use (\ref{gen}) to identify the integral of $P_{out}$ over that interval as 
the energy increment and hence $\la P_{out} \ra$, if it is a constant, as the output power. 
For large values of $\Dt$,
substituting $\bout(t) = e^{-i \ome_p t / 2} \tbout(t)$, 
expressing $\tbout(t)$ for
$t\in (t_j, t_{j+1})$ in terms of $B_k$ by the inverse cosine transform, 
and using (\ref{Bdiag}), we obtain
\be
\la P_{out} \ra \approx \half \ome_p \frac{dN_{out}}{dt} = 
\frac{\ome_p \Gam f^2}{4 \lam_1 \lam_2} \, .
\label{Pout2}
\ee
This can be interpreted as each output quantum carrying 
an average energy of $\half \ome_p$.

The amount (\ref{Pout2}) can be compared to the power 
supplied to the paramp by the drive.
By Noether's theorem, the only contribution to that power comes from the time-dependent
second term in (\ref{ham}) and equals
\be
P_{par} = \half \ome_p f \la  (\ta^\dagger)^2 + \ta^2  \ra \, .
\label{Ppar}
\ee
At large times, the paramp coherences
appearing in (\ref{Ppar}) can be computed from (\ref{corX2}) with the help of 
expressions similar to (\ref{tbout2}), except now for the paramp operators,
\be
\ta(t) = \half \cN^{1/2} \left[ X_1(t) e^{-i\phi} + i X_2(t) e^{i\phi} \right] \, .
\label{taout}
\ee
The final result matches (\ref{Pout2}). This stands to reason: 
in the steady state that the paramp reaches at large times, it
directs all the power received from the drive into the output line.

\section{Transfer of entanglement} \label{sec:mech}

Our computation of the entropy flow has highlighted the role of the off-diagonal
coherences (\ref{Boff}) in the definition of the quantum state. These coherences
develop over timescales set by the dynamics of the paramp, 
to reach the values that are just right to make the entropy flow rate vanish. 
The largest of these timescales is associated with the time constant
\be
\tau = 1 / |\lam_1| \, .
\label{tau}
\ee
Note that, as a consequence of the 
integration in (\ref{Zjk}), the coherences (\ref{Boff}) characterize 
two-point correlations of the original operators
$\bout(t)$, $\bout^\dagger(t)$ at different times, with the characteristic separation 
between these times again
determined by the timescale (\ref{tau}).
Note also that this timescale 
can be made arbitrarily large by bringing $\lam_1$ close to zero, i.e.,
the paramp close to the threshold.

For the system at hand, it is easy to come up with a qualitative argument that directly 
connects the growth of the off-diagonal coherences to the quenching of the entropy flow.
Indeed, let us switch to the 
Schr\"{o}dinger picture and suppose that, by some time $t_1$, the 
drive has produced an entangled state of the paramp mode, of the form
\be
|\Psi \ra  = [c_1 + c_2 (a^\dagger)^2 ] |0 \ra \, ,
\label{Psi}
\ee
where $|0\ra$ is the vacuum of all operators, and $c_1$ and $c_2$ are nonzero complex 
coefficients. Let us assume, for the purpose of this qualitative argument, that the
subsequent evolution of this state, until some time $t_2$, is due entirely to free 
streaming of quanta out of the paramp and can be lumped into the
action of the unitary
\be
U = \exp \left[ \frac{\pi}{2} (b^\dagger a - a^\dagger b) \right] ,
\label{U}
\ee
where $b$ is one of operators of the bath. 
This has the effect of
replacing $a^\dagger$ in (\ref{Psi}) with $b^\dagger$. As a result, there is now an
off-diagonal coherence $\la b^2 \ra \neq 0$
in the bath. On the other hand, since no entanglement 
between the output and paramp modes is left in the final state, no output entropy has 
been produced. We can say that entanglement has been transferred from the paramp mode
to the output.

If generic, the relation between off-diagonal coherences and the entropy flow would imply 
that computations that neglect the former cannot be counted upon to correctly predict 
the latter. 
One example where off-diagonal coherences
are notably absent, is Hawking's calculation of radiation from a black hole 
\cite{Hawking,Hawking:1976}. A number of authors had argued that small corrections to 
Hawking's result can accumulate and produce a result for the entropy of the radiation 
that decreases at large times; for a recent review of the work
in this direction, see Ref.~\cite{Almheiri&al}. 

From the present standpoint, it is
of interest to look for a possible mechanism of entanglement transfer in the
black-hole case. We then have to consider products of states like (\ref{Psi}) referring
to different radiation modes, for example,
\be
|\Psi \ra = \half (1 + a_1^\dagger b_1^\dagger) (1 + a_2^\dagger b_2^\dagger) |0\ra \, .
\label{Psi2}
\ee
The photons $a_1$, $a_2$ fall into the black hole, while the photons $b_1$, $b_2$
travel out; in the context of the black-hole information problem, the latter represent
the early and late radiation, respectively. The key difference between the paramp and
a black hole is that, while in the former case the drive (in the present model) is
classical and so cannot entangle with either the paramp or the bath, 
a black hole is expected to be a quantum system in its own right. As a
consequence, the evolution of the photons that have fallen in
does not have to be unitary by itself: it can involve projections (``measurements'') 
representing the effect of entanglement of these photons
with the black hole's degrees of freedom. Thus, for example, a projection
of (\ref{Psi2}) onto the state
\be
|\Phi\ra_a = \frac{1}{\sqrt{2}} (1 + a_1^\dagger a_2^\dagger) |0\ra_a \, ,
\label{Phi}
\ee
where $|0\ra_a$ is the vacuum of the operators $a$, results in the state
\be
|\mathrm{X} \ra_b = \frac{1}{\sqrt{2}} (1 + b_1^\dagger b_2^\dagger) |0\ra_b 
\label{Xstate}
\ee
for the outgoing radiation. We again observe transfer of entanglement to 
the output. In fact, one may notice that this
version of the transfer is essentially identical to the entanglement 
swapping protocol in quantum optics \cite{Zukowski&al}.

As long as the precise mechanism responsible for a projection onto a state like
(\ref{Phi}) is left unspecified, a discussion along these lines
will necessarily remain schematic. One may be able to obtain more definitive results,
however, by considering those aspects of the information 
problem that are less sensitive to the black hole's
inner workings. For instance, it has been argued that a large black hole initially 
releases information into radiation very slowly, possibly at a rate that vanishes
exponentially as a function of the black hole's thermodynamic entropy \cite{Page}. 
Assuming there is indeed a generic connection between the evolution of the entropy flow 
and the development of off-diagonal coherences in radiation, one may consider using
experimental measurements of the latter to learn about the rate of the former. In the 
next section, we present a proposal for carrying out such a measurement in an ensemble
of cold atoms.

\section{Conclusion} \label{sec:concl}

In this work, we have attempted to construct thermodynamics of quantum circuits by 
discretizing signals in time, in analogy with discretizing systems in space in the
standard thermodynamics. Our main focus has been the question of whether there is a
useful definition of entropy. We considered in detail the case of a degenerate paramp driven
by a classical pump and coupled to a zero-temperature (vacuum) 
Markovian bath. In this case, discretization is in fact required if one wishes to
resolve the delta-function correlator characterizing the Markovian noise. 

Focusing on the behavior of the system at large times,
we have presented a version of discretization that
results in a well-defined Gaussian quantum state.
Our main result is that the flux of the output entanglement entropy 
in this state vanishes. 
On the one hand, this is consistent with the interpretation of the discretization-based
definition of entropy as information, resulting in the picture where the entropy 
flow stops once all the information about the initial state of the paramp has been released.
On the other hand, the vanishing  entropy flux stands in contrast to the fluxes
of the energy and number of output quanta, both of which are nonzero. In other words, the
output signal in this case
can be viewed as a stream of multimode squeezed vacua (with the modes
corresponding to the components of the cosine transform), with small residual correlations
among the states of the stream.  
We have highlighted the role played by entanglement transfer to the output 
in the quenching of the entropy flow and noted possible relevance of this mechanism
to the black-hole information problem. 

It may be interesting to extend the present study to the case of a 
nondegenerate paramp---the system where the drive couples two types of paramp operators,
$a_1$ and $a_2$, each interacting with its own bath. The result parallel to the one
presented here would be quenching of the entropy flux between the 
paramp and the totality of the bath modes, including both baths.
That would be consistent, in our present view, with development of the off-diagonal coherences
between the output operators corresponding to different baths; such coherences are indeed 
well known to form in this system 
(see, for example, Ref.~\cite{Clerk&al}).

The connection between the entropy flow and development of the off-diagonal coherences
in the output
suggested by the present study leads us to propose a method to experimentally measure 
the rates at which high-entropy subsystems release information
about their initial states to the environment. The key point is that, while
the entanglement entropy (as a measure of the released
information) may be prohibitively difficult to measure, coherence of the outgoing 
radiation does not have to be so. 
Systems allowing local control of the interactions, such as cold gases in optical 
lattices seem especially promising in this regard. 
Let $a_1$ and $a_2$ represent two oscillation modes of a Bose-Einstein condensate
of atoms in a spatial region. A time-dependent modulation of the coupling between the atoms 
there will give rise to a pair-production Hamiltonian $f(t) a_1^\dagger a_2^\dagger$ 
(plus the Hermitian conjugate), analogous to the pump term in (\ref{ham}). 
While a periodic $f(t)$ would be the closest to 
the case considered here, other forms of time dependence may in practice do as well.
Consider for instance a gas quenched to a strong 
interaction at the center of a box trap, while remaining weakly interacting in the outer
regions. The off-diagonal coherences of quasiparticles (Bogoliubov phonons) in the 
weakly coupled regions are measurable quantities; for a uniform quench in a 
weakly coupled gas, they in fact have been measured \cite{Chen&al}.
Provided these coherences are not destroyed too quickly by collisions, 
one may be able to study their evolution
after the quench and thereby extract the timescale on which information is released 
from the central region.

\begin{acknowledgments}
The author is grateful to Chen-Lung Hung for a discussion.
This work was supported in part by the DOE QuantISED program through a theory consortium 
at Fermilab.
\end{acknowledgments}


\begin{thebibliography}{99}
\bibitem{LL} L. D. Landau and E. M. Lifshitz, {\it Statistical Physics. Part 1},
3rd edition (Butterworth-Heinemann, 1980).
\bibitem{Gabor} D. Gabor, Theory of communication,
Journal I. E. E. - Part III: Radio and Communication Engineering {\bf 93}, 429 (1946).
\bibitem{modes_of_univ} J. Gea-Banacloche, N. Lu, L. M. Pedrotti, S. Prasad, M. O.
Scully, and K. W\'{o}dkiewicz, 
Treatment of the spectrum of squeezing based on the modes of the universe. 
I. Theory and a physical picture,
Phys. Rev. A {\bf 41}, 369 (1990).
\bibitem{Clerk&al} A. A. Clerk, M. H. Devoret, S. M. Girvin, F. Marquardt, and
R. J. Schoelkopf,
Introduction to quantum noise, measurement, and amplification,
Rev. Mod. Phys. {\bf 82}, 1155 (2010) [arXiv:0810.4729].
\bibitem{Collett&Gardiner} M. J. Collett and C. W. Gardiner,
Squeezing of intracavity and traveling-wave light fields produced in parametric 
amplification,
Phys. Rev. A {\bf 30}, 1386 (1984).
\bibitem{Gardiner&Collett} C. W. Gardiner and M. J. Collett,
Input and output in damped quantum systems: Quantum stochastic differential equations 
and the master equation,
Phys. Rev. A {\bf 31}, 3761 (1985).
\bibitem{Williamson} J. Williamson, On the Algebraic Problem Concerning the Normal Forms 
of Linear Dynamical Systems,
Am. J. Math. {\bf 58}, 141 (1936).
\bibitem{Simon&al} R. Simon, N. Mukunda, and B. Dutta,
Quantum-noise matrix for multimode systems: U(n) invariance, squeezing, and normal forms,
Phys. Rev. A {\bf 49}, 1567 (1994).
\bibitem{Hawking} S. W. Hawking, Particle Creation by Black Holes,
Commun. Math. Phys. {\bf 43}, 199 (1975) [Erratum: {\it ibid.} {\bf 46}, 206 (1976)].
\bibitem{Hawking:1976} S. W. Hawking, Breakdown of predictability in gravitational 
collapse,
Phys. Rev. D {\bf 14}, 2460 (1976).
\bibitem{Almheiri&al} 
A. Almheiri, T. Hartman, J. Maldacena, E. Shaghoulian, and A. Tajdini,
The entropy of Hawking radiation,
Rev. Mod. Phys. {\bf 93}, 035002 (2021) [arXiv:2006.06872].
\bibitem{Zukowski&al} M. \.{Z}ukowski, A. Zeilinger, M. A. Horne, and A. K. Ekert,
``Event-Ready-Detectors" Bell Experiment via Entanglement Swapping,
Phys. Rev. Lett. {\bf 71}, 4287 (1993).
\bibitem{Page} D. N. Page, Information in Black Hole Radiation, 
Phys. Rev. Lett. {\bf 71}, 3743 (1993) [arXiv:hep-th/9306083].
\bibitem{Genoni&al} M. G. Genoni, L. Lami, and A. Serafini,
Conditional and unconditional Gaussian quantum dynamics,
Contemporary Physics {\bf 57}, 331 (2016) [arXiv:1607.02619].
\bibitem{HBosons} S. Khlebnikov and M. Kruczenski, 
Locality, entanglement, and thermalization of isolated quantum systems.
Phys. Rev. E {\bf 90}, 050101(R) (2014); Thermalization of isolated quantum systems,
arXiv:1312.4612.
\bibitem{fheat} S. Khlebnikov and A. Sheoran, 
The first heat: production of entanglement entropy in the early universe,
JHEP {\bf 11}, 157 (2019) [arXiv:1907.00487].
\bibitem{Gaudin} M. Gaudin,
Une démonstration simplifiée du théorème de Wick en mécanique statistique,
Nucl. Phys. {\bf 15}, 89 (1960).
\bibitem{Chen&al} C.-A. Chen, S. Khlebnikov, and C.-L. Hung, 
Observation of Quasiparticle Pair Production and Quantum Entanglement in Atomic 
Quantum Gases Quenched to an Attractive Interaction,
Phys. Rev. Lett. {\bf 127}, 060404 (2021) [arXiv:2102.11215].

\end{thebibliography}
\end{document}